\documentclass[a4paper]{jpconf}

\usepackage{amsmath}
\usepackage{amsfonts}
\usepackage{amsthm}
\usepackage{amstext}
\usepackage{amssymb}
\usepackage{amsbsy}
\usepackage{amscd}
\usepackage{graphicx}
\usepackage{iopams}
\usepackage[tight]{subfigure}
\usepackage[modulo]{lineno}
\usepackage{color}
\usepackage{type1cm}
\usepackage{cite}

\begin{document}
\title{The effects of nonlinear wave propagation on the stability of inertial cavitation}
\author{D~Sinden, E~Stride \& N~Saffari}
\address{Mechanical Engineering, Roberts Building, Torrington Place, London, WC1E 7JE, United Kingdom}
\ead{d.sinden@ucl.ac.uk}
\begin{abstract}
In the context of forecasting temperature and pressure fields in high-intensity focussed ultrasound, the accuracy of predictive models is critical for the safety and efficacy of treatment. In such fields inertial cavitation is often observed. Classically, estimations of cavitation thresholds have been based on the assumption that the incident wave at the surface of a bubble was the same as in the far-field, neglecting the effect of nonlinear wave propagation. By modelling the incident wave as a solution to Burgers' equation using weak shock theory, the effects of nonlinear wave propagation on inertial cavitation are investigated using both numerical and analytical techniques. From radius-time curves for a single bubble, it is observed that there is a reduction in the maximum size of a bubble undergoing inertial cavitation and that the inertial collapse occurs earlier in contrast with the classical case. Corresponding stability thresholds for a bubble whose initial radius is slightly below the critical Blake radius are calculated. Bifurcation diagrams and frequency-response curves are presented associated with the loss of stability. The consequences and physical implications of the results are discussed with respect to the classical results.
\end{abstract}
\section{Introduction} \label{sec:introduction}
\par
For materials with a nonlinear stress-strain relationship, such as tissue~\cite{picinbono2001nae}, the point of maximum compression for a wave may propagate faster than the point of maximum rarefaction, leading to a distortion in the wave profile and the redistribution of energy from the fundamental harmonic frequency to higher harmonics. The pressures associated with therapeutic high intensity focused ultrasound may be high enough for the effects of nonlinear wave propagation to be significant~\cite{muir1980pna}. As high frequency components are absorbed more easily than those with lower frequencies, nonlinear wave propagation contributes to increased absorption, enhanced heating and a subsequent shift in the focal point of targeted ultrasonic beam, potentially damaging healthy tissue. As well as providing greater predictive accuracy, knowledge of nonlinear wave propagation will enable increased accuracy in calibration using the received signal generated by bubble oscillations~\cite{humphrey2000npu} and thus greater accuracy in treatment procedures. In this paper the implications of nonlinear wave propagation on inertial cavitation are investigated and stability criteria re-derived in an attempt to reconcile theory and experiment.
\par
The Rayleigh-Plesset equation~\cite{lauterborn1976nin} is a nonlinear equation which determines the size of a spherical bubble subject to a varying pressure field. Wave propagation from the far-field to the surface of the bubble is generally assumed to be linear, yet this does not correspond with experimental observations in the context of high intensity focused ultrasound (HIFU). Moss~\cite{moss1997upd} attempted to incorporate the effects of boundary conditions and global compressibility/local incompressibility into the Rayleigh-Plesset equation but only considered linear wave propagation.  The equation derived was identical to the classical Rayleigh-Plesset equation if the far-field pressure is replaced by an attenuated pressure at the bubble surface. In this paper the effect of distortion rather than attenuation of the wave profile and its effect on the stability of oscillations will be investigated.
\par
Lauterborn and co-workers~\cite{lauterborn1976nin,lauterborn1981src} and Smerka~\cite{smereka1987rcb} showed experimentally and numerically that bubble oscillations undergo a sequence of period-doubling bifurcations leading to unstable quasi-periodic (chaotic) oscillations. The period doubling route to chaos occurs through a succession of saddle-node bifurcations of subharmonic periodic orbits. Homoclinic bifurcations are the limit of a countable sequence of subharmonic saddle-node bifurcations and thus provide an insight into the parameters at which unbounded growth may begin to occur.
\par
Mel'nikov's method provides a measure of the distance between the stable and unstable manifolds of a periodically perturbed system. If the manifolds intersect transversely once, they will intersect transversely infinitely many more times. The transverse intersections can be represented by Smale horseshoes, which through the Smale-Birkhoff theorem give an elegant description of sensitivity to initial conditions and the resulting chaotic oscillations~\cite{guckenheimer1983nod}. The first application of Mel'nikov's method to cavitation was performed by Chang and Chen~\cite{chang1986ggb} on the effect of viscosity on the Hamiltonian structure. Harkin~\cite{harkin1999acs} also performs Mel'nikov analysis on bubbles whose initial radius is slightly smaller than Blake critical radius. Using matched perturbation analysis, Harkin derives a second order normal form for the Rayleigh-Plesset equation. The normal form is a damped driven oscillator. An escape velocity, like the static Blake criterion, provides an upper bound for when unbounded growth will occur, whereas Mel'nikov's method provides a lower bound for when the transition to chaos and unbounded growth may occur. The fate of bubbles whose initial conditions lie in the intermediate region between the Mel'nikov and Blake thresholds can be computed by a transport-type processes~\cite{kang1990bdt}. A Bernoulli shift map on two symbols has already been constructed numerically from a bifurcation diagram for the full Rayleigh-Plesset equation~\cite{simon2002pot} without explicit inference to sensitivity to initial conditions. In each application of Mel'nikov method the incident pressure wave was sinusoidal~\cite{smereka1987rcb,chang1986ggb,harkin1999acs,kang1990bdt,simon2002pot,szeri1991oco}. In this paper the analysis is performed for nonlinear waves. 
\par
The outline of this paper is as follows, in section~\ref{sec:nonlinearity} nonlinear wave propagation is considered and wave profiles derived. Then in section~\ref{sec:Rayleigh_Plesset} the Rayleigh-Plesset equation is introduced and the effects of nonlinear wave propagation investigated. In section~\ref{sec:melnikov} Mel'nikov analysis is performed for nonlinear wave propagation, providing an improved measure of the values at which quasi-periodic oscillation and unbounded growth may occur. Finally, conclusions and implications are discussed in section~\ref{sec:conclusions}.
\section{Nonlinear Wave Propagation} \label{sec:nonlinearity}
There is no universally accepted system of partial differential equations for the modelling of ultrasound propagation in biological tissue. Perhaps the best known is the Khokhlov-Zabolotskaya-Kuznetsov (KZK) equation. The KZK equation is a parabolic wave equation which includes the effects of diffraction, absorption and nonlinearity of the directed beams~\cite{humphrey2000npu}. The KZK equation for an axi-symmetric beam which propagates in the~$r$ direction is written in terms of the acoustic pressure~$p\left(r,t\right)$ as
\begin{equation}
\dfrac{\partial^{2} p}{\partial r \partial t^{\prime} } = \dfrac{c_{0}^{}}{2} \nabla_{\perp}^{2} p + \dfrac{D}{2c_{0}^{3}} \dfrac{\partial^{3}p}{\partial t^{\prime 3}} + \dfrac{\beta}{2 \rho_{}^{} c_{0}^{3} } \dfrac{\partial^{2} p^{2} }{\partial t^{\prime 2}} \label{eq:kzk0}
\end{equation}
where~$c_{0}$ is the initial wave speed,~$\rho_{}$ is the density of the medium,~$\beta$ is the standard nonlinearity coefficient given by~${\beta=1+B\slash 2 A}$, where~$B \slash A$ is the standard nonlinearity parameter,~$D$ is the sound diffusivity and~${t^{\prime}=t - r \slash c_{0}^{}}$ is the retarded time variable. The Laplacian is taken with respect to the transverse coordinates. The sound diffusivity is given by
\begin{equation}
D = \dfrac{1}{\rho_{}} \left( \mu_{b} + 4\mu_{s} \slash 3 + \kappa\left( 1 \slash c_{v}- 1\slash c_{p} \right) \right) \nonumber
\end{equation}
where~$\mu_{b}$ and~$\mu_{s}$ are the bulk and shear viscosity respectively,~${1\le\kappa\le\gamma}$ the polytropic exponent, and~$c_{v}$ and~$c_{p}$ are the specific heats at constant volume and pressure respectively. If~${\kappa=1}$ the system is isothermal, if~${\kappa=\gamma \equiv c_{p}\slash c_{v}}$ the system is adiabatic. In the study of cavitation it will be assumed that the distance from the surface of the bubble to the shock front is a fixed length. The equation~\eqref{eq:kzk0} can be integrated to give
\begin{equation}
\dfrac{\partial p}{\partial t^{\prime} } = \dfrac{c_{0}^{}}{2} \int_{-\infty}^{r} \nabla_{\perp}^{2} p\left(s,t\right) \, \mathrm{d}s + \dfrac{D}{2c_{0}^{3}} \dfrac{\partial^{2}p}{\partial r^{2}} + \dfrac{\beta}{2 \rho_{}^{} c_{0}^{3} } \dfrac{\partial^{} p^{2} }{\partial r}. \label{eq:kzk}
\end{equation}
Discarding the effects of diffraction Burgers', or alternatively the Burgers-Hopf, equation is recovered
\begin{align}
\dfrac{\partial p}{\partial r} & = \dfrac{D}{2c_{0}^{3}} \dfrac{\partial^{2} p}{\partial t^{\prime 2} } + \dfrac{\beta}{2\rho_{}c_{0}^{3}} \dfrac{\partial p^{2}}{\partial t^{\prime}}. \label{eq:burgers_hopf}
\end{align}
If~${D=0}$ the fluid is inviscid, and the so-called lossless Burgers equation is recovered.
\par
From a given driving pressure,~$f\left(t\right)$, the incident pressure wave,~$p\left(t\right)$, can be expressed using weak shock theory. The location of a shock is determined by the Rankine-Hugoniot relation defining the conservation of flux. The areas enclosed by the multi-valued solution to the left and right of the shock are equal. Thus, by this symmetry, the shock is positioned at the zero of the linear solution. For a sinusoidal driving pressure of magnitude~$P$ and frequency~$\omega$, i.e.~${f\left(t\right)=P\sin\left(\omega t\right)}$, a Fourier expansion of the solution to the lossless Burgers equation yields the Bessel-Fubini solution
\begin{align}
p\left(t\right) & = \dfrac{2P}{r_{s}}\sum_{n=1}^{\infty} \dfrac{1}{n} J_{n}\left( n r_{s} \right) \sin \left( \omega_{n} t \right), \label{eq:bessel_fubini}
\end{align}
where~${\omega_{n}=n \omega}$,~$J_{n}$ is an~$n^{\mathrm{th}}$-order Bessel functions of the first kind and $r_{s}$ is the normalised distance to a shock given by
\begin{equation}
r_{s} = \dfrac{r}{r_{c}} \quad \mbox{where} \quad r_{c} = \dfrac{ \rho c_{0}^{3} }{ \beta \omega P } \nonumber 
\end{equation}
is the location of the shock.
\par
Beyond the shock, weak shock theory can once again be employed to find a solution, however, the resulting analytical solution takes an integral form. An asymptotic solution, valid for~${r_{s}>3}$ is 
\begin{equation}
p = \dfrac{2P}{1+r_{s}}\sum_{n=1}^{\infty} \dfrac{1}{n} \sin\left(\omega_{n} t^{\prime} \right)
\end{equation}
which can be expressed in the time domain as 
\begin{equation}
p \slash P = \left\{ \begin{array}{ll}
- \dfrac{\omega t^{\prime} + \pi}{1+r_{s}}  \quad \mbox{for} \quad -& \!\!\!\! \pi < \omega t^{\prime} < 0, \\
- \dfrac{\omega t^{\prime} - \pi}{1+r_{s}}\quad \mbox{for} \quad & \!\!\!\! 0 < \omega t^{\prime} < \pi.
\end{array} \right. \label{eq:saw}
\end{equation}
\par
A general solution to the Burgers'-Hopf equation for a sinusoidal driving pressure, derived by transforming the nonlinear equation into a linear diffusion equation via the Hopf-Cole transformation, is given by
\begin{equation}
p = - \dfrac{4 P}{\kappa} \dfrac{\sum_{n=1}^{\infty} n \left(-1\right)^{n} I_{n}\left(\Gamma\slash{2}\right) e^{-n^{2}r_{s} \slash \kappa } \sin \left(\omega_{n}t^{\prime}\right)}{I_{0}\left(\Gamma\slash{2}\right) + 2 \sum_{n=1}^{\infty} \left(-1\right)^{n} I_{n}\left(\Gamma\slash{2}\right) e^{-n^{2}r_{s} \slash \kappa } \cos\left(\omega_{n}t^{\prime}\right) }
\end{equation}
where~$I_{n}$ is an~$n^{\mathrm{th}}$-order modified Bessel functions of the first kind and~$\Gamma$ is the Gol'dberg number which relates the importance of nonlinear effects against dissipative effects by
\begin{equation}
\Gamma = \dfrac{2\beta P}{D \rho \omega }. \nonumber 
\end{equation}
If~${\Gamma \gg 1}$ then the relative effects of nonlinearity are strong, if~${\Gamma < 1}$ the relative effects of nonlinearity are weak. 
\par
Unlike~\eqref{eq:bessel_fubini}, for strong nonlinearity, i.e.~${\Gamma \gg 1}$, the solution converges slowly but far from the shock, i.e.~${r_{s} > 3}$, asymptotic analysis yields the more usable formulation
\begin{equation}
p = \dfrac{2 P}{\Gamma} \sum_{n=1}^{\infty} \dfrac{\sin \left(\omega_{n}t^{\prime}\right)}{\sinh\left( n\left(1+r_{s} \right)\slash \Gamma \right)}  \label{eq:fay}
\end{equation}
called the Fay solution. The Fay solution can be expressed as the periodic function
\begin{equation}
p = \dfrac{P}{1+r_{s}} \left( - \omega t^{\prime} + \pi \tanh\left( \dfrac{\pi \Gamma \omega t^{\prime}}{2\left(1+r_{s}\right)} \right) \right) \quad -\pi \le \omega t^{\prime} \le \pi.
\end{equation}
In the lossless limit as viscosity vanishes, that is as~${\Gamma \rightarrow \infty}$, the Fay solution recovers the Fourier expression for a sawtooth function~\eqref{eq:saw}.  The Fubini~\eqref{eq:bessel_fubini} and Fay~\eqref{eq:fay} solutions may at first seem incompatible but each holds in a different region of the flow; the Fubini solution close to the source as shocks begin to form and the Fay solution far from the source as shocks begin to decay.  Blackstock~\cite{blackstock1966cbf} shows that the true solution to the lossless Burgers equation is simply the sum of the two solutions~\eqref{eq:bessel_fubini} and~\eqref{eq:saw}.  In the near-field the Fubini solution is dominant, in the far-field the sawtooth solution is dominant. In~\cite{mitome1989esf} exact solutions of Burgers' equation from the Cole-Hopf transformation were computed numerically and contrasted against the solutions and showing good agreement between the sets of solutions unless in the immediate neighbourhood of the shock. Furthermore, the formation of a shock will result in a negative pressure and hence is a prime site for the nucleation of cavities. Thus the forthcoming analysis will be performed either before a shock or significantly after, so that only the effect of nonlinear wave propagation on pre-existing cavities is studied. 
\par
\begin{figure}
\begin{center}
\begin{picture}(0,0)%
\includegraphics[scale=1.25]{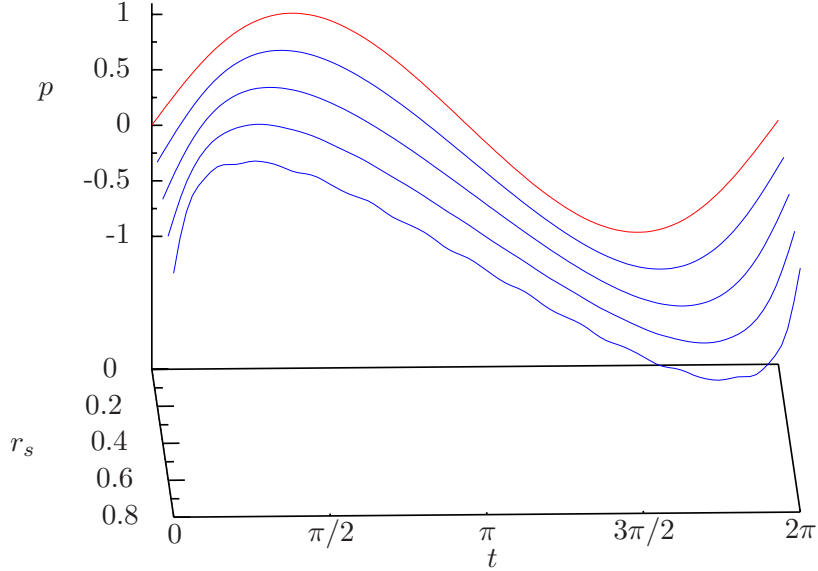} %
\end{picture}%
\begingroup
 \setlength{\unitlength}{0.0625bp}%
  \begin{picture}(7200.00,5040.00)%
      \put(1459,1955){\makebox(0,0)[r]{\strut{} 0}}%
      \put(1491,1733){\makebox(0,0)[r]{\strut{} 0.2}}%
      \put(1524,1511){\makebox(0,0)[r]{\strut{} 0.4}}%
      \put(1557,1288){\makebox(0,0)[r]{\strut{} 0.6}}%
      \put(1589,1066){\makebox(0,0)[r]{\strut{} 0.8}}%
      \put(1802,956){\makebox(0,0){\strut{}$0$}}%
      \put(2736,964){\makebox(0,0){\strut{}$\pi \slash 2$}}%
      \put(3670,972){\makebox(0,0){\strut{}$\pi$}}%
      \put(4604,980){\makebox(0,0){\strut{}$3\pi \slash 2$}}%
      \put(5538,987){\makebox(0,0){\strut{}$2\pi$}}%
      \put(1540,2758){\makebox(0,0)[r]{\strut{}-1}}%
      \put(1540,3092){\makebox(0,0)[r]{\strut{}-0.5}}%
      \put(1540,3426){\makebox(0,0)[r]{\strut{} 0}}%
      \put(1540,3760){\makebox(0,0)[r]{\strut{} 0.5}}%
      \put(1540,4093){\makebox(0,0)[r]{\strut{} 1}}%
      \put(897,1506){\makebox(0,0){\strut{}$r_s$}}%
      \put(3697,812){\makebox(0,0){\strut{}$t$}}%
      \put(1034,3648){\makebox(0,0){\strut{}$p$}}%
\end{picture}%
\endgroup
\caption[]{Profiles of incident waves for nonlinear wave propagation. The nonlinear wave is modeled by the first twenty terms of the Bessel-Fubini solution. \label{fig:waves}  }
\end{center}
\end{figure}

Figure~\ref{fig:waves} contrasts the nonlinear wave profile and the linear wave profile illustrating the distortion due to nonlinear propagation of waves with equal amplitude. Throughout this paper comparisons between linear and nonlinear waves of equal amplitude are studied.
%
%
\section{Rayleigh-Plesset Equation} \label{sec:Rayleigh_Plesset}
The Rayleigh-Plesset equation is an ordinary differential equation which models the oscillations of a spherical bubble of radius~$R$
\begin{equation}
\rho\left( R\ddot{R} + \dfrac{3}{2}\dot{R}^{2} \right) = p_{g}\left(R\right) + p_{v} - p_{\infty} + p\left(t\right) + \dfrac{2\sigma}{R} + \dfrac{4\mu\dot{R}}{R} \label{eq:RP}
\end{equation} 
where~$p_{g}$ is the internal pressure of the gas in the bubble given by the hardcore van der Waals relationship
\begin{equation}
p_{g}\left(R\left(t\right)\right) = \left( p_{\infty} - p_{v} + \dfrac{2\sigma}{R_{0}} \right)\left( \dfrac{R_{0}^{3}-a_{}^{3}}{R\left(t\right)_{}^{3}-a_{}^{3}} \right)^{\kappa} \
\end{equation}
with~$R_{0}$ the equilibrium radius,~$a$ the van der Waals hard-core radius,~$p_{v}$ the vapour pressure,~$p_{\infty}$ is the ambient pressure,~$\sigma$ is the surface tension and~$\mu$ viscosity.  The gas is assumed to be ideal as the internal pressure is a function of the bubble radius only. The forcing pressure will take the form
\begin{equation}
p\left(t\right) = \sum_{n=1}^{\infty} P_{n} \sin \left( \omega_{n} \left(t + t_{0}\right) \right) \nonumber
\end{equation}
where~$P_{n}$ are the Fourier terms of a solution to a nonlinear wave equation, such as those given by the Bessel-Fubini~\eqref{eq:bessel_fubini} or the Fay~\eqref{eq:fay} solutions. The Rayleigh-Plesset equation can be derived by balancing the energy supplied to the bubble by the incident pressure and the surround fluid and the kinetic energy of the bubble oscillations~\cite{leighton1994ab}.
\par
\begin{figure}
\begin{center}
\begin{picture}(0,0)%
\includegraphics[scale=1.0]{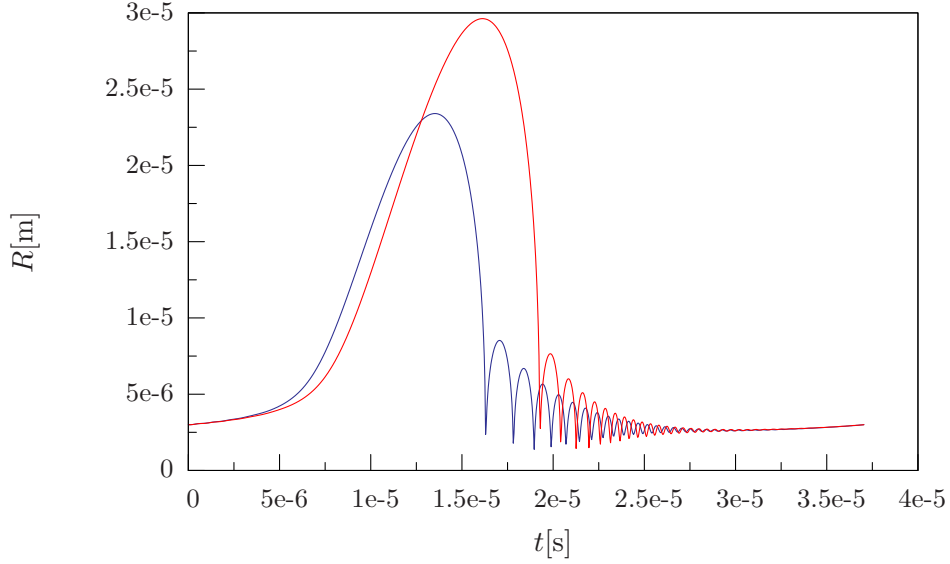}%
\end{picture}%
\begingroup
\setlength{\unitlength}{0.0200bp}%
\begin{picture}(18000,10800)(0,0)%
  \put(3300,1650){\makebox(0,0)[r]{\strut{} \small 0}}%
  \put(3300,3083){\makebox(0,0)[r]{\strut{} \small 5e-6}}%
  \put(3300,4517){\makebox(0,0)[r]{\strut{} \small 1e-5}}%
  \put(3300,5950){\makebox(0,0)[r]{\strut{} \small 1.5e-5}}%
  \put(3300,7383){\makebox(0,0)[r]{\strut{} \small 2e-5}}%
  \put(3300,8817){\makebox(0,0)[r]{\strut{} \small 2.5e-5}}%
  \put(3300,10250){\makebox(0,0)[r]{\strut{} \small 3e-5}}%
  \put(3575,1100){\makebox(0,0){\strut{} \small 0}}%
  \put(5275,1100){\makebox(0,0){\strut{} \small 5e-6}}%
  \put(6975,1100){\makebox(0,0){\strut{} \small 1e-5}}%
  \put(8675,1100){\makebox(0,0){\strut{} \small 1.5e-5}}%
  \put(10375,1100){\makebox(0,0){\strut{} \small 2e-5}}%
  \put(12075,1100){\makebox(0,0){\strut{} \small 2.5e-5}}%
  \put(13775,1100){\makebox(0,0){\strut{} \small 3e-5}}%
  \put(15475,1100){\makebox(0,0){\strut{} \small 3.5e-5}}%
  \put(17175,1100){\makebox(0,0){\strut{} \small 4e-5}}%
  \put(550,5950){\rotatebox{90}{\makebox(0,0){\strut{}$R$[m]}}}%
  \put(10375,275){\makebox(0,0){\strut{}$t$[s]}}%
\end{picture}%
\endgroup
\caption[short]{Contrasting profiles of inertial cavitation for linear (red) and nonlinear (blue) wave propagation modeled by the first ten terms of the Bessel-Fubini solution.~${R_{0}=0.6\mu}$m, subject to a wave with frequency~${\omega=2\pi \cdot 26}$kHz, amplitude of driving pressure~${P=1.36}$atm, normalised distance to the shock~${r_{s}=1 \slash 20}$ ambient pressure~$p_{\infty}=1$atm, surface tension~${\sigma =0.073}$kgm$^{-2}$, viscosity~${\mu =10^{-3}}$kgm$^{-1}$s$^{-1}$, density~${\rho=1000}$kgm$^{-3}$, polytropic exponent~${\kappa =5\slash 3}$ and hardcore radius~${a=R_{0}\slash 8.85}$.} \label{fig:contrast1}
\end{center}
\end{figure}

Figure~\ref{fig:contrast1} contrasts the effects of nonlinear and linear wave propagation. It is clear that for nonlinear wave propagation inertial cavitation occurs before and at a smaller maximum radius than for linearly propagated waves. This has two significant effects, firstly as the collapse occurs at a small maximum radius there is a diminished chance of shape instability~\cite{brenner1995bso} and secondly, as the collapse occurs earlier more after-bounces can occurs so that the bubble returns to a stable initial radius before the next cycle. Note that the period of the after-bounces is the same for the nonlinear and linear wave since the after-bounces occur at the natural frequency of the Rayleigh-Plesset equation which is independent of the applied pressure.  
\section{Mel'nikov Analysis} \label{sec:melnikov}
\par
Considering an unforced bubble and, discarding the effects of viscosity and the derivative of the gas pressure, let~${R\left(t\right)=R_{0}\left(1+x\right)}$.  To first order, the nondimensional Rayleigh-Plesset equation yields the simple harmonic equation
\begin{equation}
\ddot{x} + \omega_{0}^{} x = 0 \quad \mbox{where} \quad \omega_{0}^{2} = \dfrac{1}{\rho}\left( \dfrac{ 3 \kappa \left(p_{\infty}^{}-p_{v}\right) R_{0}^{} }{ R_{0}^{3} - a_{}^{3} } + 2\sigma \left( \dfrac{3\kappa}{R_{0}^{3}-a_{}^{3}} -\dfrac{1}{R_{0}^{3}} \right) \right).
\end{equation}
The natural frequency~${\omega_{0}^{}}$ is used to nondimensionalise time by~${\tau=\omega_{0}^{} t}$ so that the new time-like variable will be a function of the perturbation parameter. Note that when~${a=0}$ and~${\kappa=1}$ then the frequency given by Harkin~\cite[Eq.~(12)]{harkin1999acs} is recovered. In this section Harkin's analysis is followed as it gives an analytical expression for the transverse intersections which allows for comparison between linear and nonlinear wave propagation. 
\par
The static Blake threshold pressure is the point at which the internal pressure ${p_{v}+p_{g}}$ is equal to the external pressure ${p_{l}+2\sigma\slash R}$, thus for internal pressures larger than this threshold unbounded growth will occur. The equilibrium pressure exerted on the bubble surface by the liquid,~$p_{l}$, is given by
\begin{equation}
p_{l} = p_{g} + p_{v} - \dfrac{2\sigma}{R}. \label{eq:pl}
\end{equation}
Now perturb the equilibrium radius by ${R\left(t\right)=R_{0}\left(1+\epsilon x\left(\tau\right)\right)}$ with~$\epsilon$ a small parameter given by
\begin{equation}
\epsilon = 2\left( 1- {R_{0}}\slash{R_{c}} \right) \nonumber
\end{equation}
where~$R_{c}$ is the Blake critical radius, found as the stationary solutions to the unforced Rayleigh-Plesset equation
\begin{equation}
\dfrac{3\kappa\tilde{G}R^{4}}{\left(R^{3}-a^{3}\right)^{\left(\kappa+1\right)}} = 2 \sigma \quad \mbox{where} \quad \tilde{G} = \left( p_{\infty} - p_{v} + \dfrac{2\sigma}{R_{0}}\right)\left( R_{0}^{3} - a_{}^{3} \right)^{\kappa}.
\end{equation}
Unfortunately if~${a \ne 0}$ then there is no simple analytical expression for the critical Blake radius~${R_{c}=R_{c}\left(\sigma, \kappa, a\right)}$.  In the isothermal case the critical radius can be found explicitly as the solution to the cubic equation
\begin{equation}
R^{3} - \sqrt{ \dfrac{3\tilde{G}}{2\sigma} } R^{2} - a^{3} = 0. \nonumber
\end{equation}
It is straightforward to show from the discriminant of the cubic equation when~${a \ge 0}$ and ${\tilde{G} \ge 0}$ that the equation will have one real solution~$R_{c}$ and a pair of ignorable complex conjugate solutions. In the case of no hardcore radius, i.e.~${a=0}$, then
\begin{equation}
R_{c} = \sqrt[3\kappa-1]{ \dfrac{3\kappa\tilde{G}}{2\sigma} } \label{eq:Rc}
\end{equation}
so that the critical value for the liquid pressure~\eqref{eq:pl} is then
\begin{equation}
p_{c} = p_{v} - \sqrt[3\kappa-1]{ \dfrac{\left(2\sigma\right)^{3\kappa}}{3 \kappa \tilde{G}} } \left( 1-\dfrac{1}{3\kappa}\right). \label{eq:pc}
\end{equation}
On combining~\eqref{eq:Rc} and~\eqref{eq:pc} the Blake critical radius is then given in the familiar form by
\begin{equation}
R_{c} = \dfrac{2\sigma}{\left(p_{v}-p_{c}\right)}\left( 1-\dfrac{1}{3\kappa}\right). 
\end{equation}
\par
Note that 
\begin{align}
p_{\infty}^{} - p_{v}^{} & = \dfrac{2\sigma}{3 R_{0}^{}}\left( \dfrac{R_{c}^{}}{R_{0}^{}} \right)^{2} - \dfrac{2\sigma}{R_{0}^{}} \nonumber \\
& = \dfrac{2\sigma}{3\kappa R_{0}^{}}\left(1-3\kappa + \epsilon + \dfrac{1}{2}\dfrac{3\kappa}{3\kappa-1}\epsilon_{}^{2} + \mathcal{O}\left(\epsilon_{}^{3}\right) \right),
\end{align}
and similarly
\begin{align}
p_{c} - p_{v}^{} & = \dfrac{2\sigma\left(3\kappa-1\right)}{3 \kappa R_{0}^{}}\left( 1 - \dfrac{\epsilon}{2} \right)^{-1}
\end{align}
so the critical pressure~$p_{c}$ and the ambient pressure~$p_{\infty}$ in general differ by terms~$\mathcal{O}\left(\epsilon\right)$ but when isothermal~$\mathcal{O}\left(\epsilon^{2}\right)$. Thus when the equilibrium radius is close to the critical radius and the ambient pressure is close to the critical pressure, the natural frequency can be expressed as
\begin{equation}
\omega_{0}^{2} = { \dfrac{2\sigma \epsilon\left(3\kappa-1\right)}{\rho R_{0}^{3} } }.
\end{equation}
When the perturbation is applied to the Rayleigh-Plesset equation~\eqref{eq:RP} in this regime both the~$\mathcal{O}\left(1\right)$ and~$\mathcal{O}\left(\epsilon\right)$ terms are zero \emph{if and only if the system is isothermal}, i.e.~${\kappa=1}$. Thus, when isothermal to~$\mathcal{O}\left(\epsilon^{2}\right)$ the governing equation is a Helmholtz oscillator
\begin{equation}
\ddot{x} + 2 \zeta\dot{x} + x\left(1-x\right) = \sum_{n=1}^{N} A_{n} \sin \left( \Omega_{n} \left(\tau + \tau_{0}\right) \right) ,
\end{equation}
where the overdot represents the derivative with respect to the nondimensional time~$\tau$ and
\begin{equation}
\zeta = \mu\sqrt{ \dfrac{2}{\epsilon \sigma \rho R_{0}} }, \quad A_{n} = \dfrac{P_{n} R_{0}}{2\sigma\epsilon^{2}} \quad \mbox{and} \quad \Omega_{n} = \omega_{n}\sqrt{\dfrac{\rho R_{0}^{3}}{2\sigma\epsilon}} \nonumber
\end{equation}
are constant terms of~$\mathcal{O}\left(1\right)$ when~$\epsilon$ is small for bubbles whose initial radius is of order microns driven by frequencies in the megahertz range~\cite{harkin1999acs}. When $\kappa=1$ the analysis holds to $\mathcal{O}\left(\epsilon^{2}\right)$ because thermal dissipation is assumed to be negligible. Here~$\zeta$ is the nondimensional damping, $\Omega_{n}$ are harmonics of the nondimensional frequency, $A_{n}$ the nondimensional Fourier components of the applied pressure and~$\tau_{0}$ is the phase of the incident wave. Note that the series expansion is truncated to~$N$ terms in order to disregard terms which are greater than~$\mathcal{O}\left(\epsilon^{2}\right)$. 
\par
When forcing and viscosity are rescaled as small parameters, ${\xi \mapsto \varepsilon \xi}$ and ${f \mapsto \varepsilon f}$, which both destroy the integrable Hamiltonian structure then the Mel'nikov integral can be calculated. Let the $\varepsilon$-perturbed system be given by ${\dot{\boldsymbol{x}} = \boldsymbol{f}_{0}\left(\boldsymbol{x}\right) + \varepsilon \boldsymbol{f}_{1}\left(\boldsymbol{x},\tau\right)}$ with ${\boldsymbol{x}=\left(x,y\right)^{T} = \left(x, \dot{x} \right)^{T}}$ and $\boldsymbol{f}_{1}$ an $\Omega$-periodic function. Explicitly the vector field is given by
\begin{align}
\dot{x} & = y, \nonumber \\
\dot{y} & = x \left( x - 1 \right) + \varepsilon\left( \sum_{n=1}^{N} A_{n} \sin\left(\Omega_{n}\tau\right)- 2 \xi y \right). \nonumber
\end{align} 
The unperturbed system, ${\varepsilon=0}$, admits a homoclinic orbit $\boldsymbol{\gamma}$ emanating from the saddle at~$\left(1,0\right)$ of the form
\begin{equation}
\boldsymbol{\gamma}\left(\tau\right) = \dfrac{1}{2}\left( \tanh^{2}\left(\dfrac{\tau}{2}\right)-1 , 3\tanh\left(\dfrac{\tau}{2}\right) \mathrm{sech}^{2}\left(\dfrac{\tau}{2}\right) \right). \nonumber
\end{equation}
The first order Mel'nikov integral at the homoclinic energy level~$h$ can simply be calculated using Cauchy's residue theory as
\begin{align}
\mathcal{M}_{h}^{\left(1\right)}\left(\tau_{0}\right) & = \int_{-\infty}^{+\infty} \boldsymbol{f}_{0}\left(\boldsymbol{x}\right) \wedge \boldsymbol{f}_{1}\left(\boldsymbol{x},\tau+\tau_{0}\right) \, \mathrm{d}\tau \nonumber \\
& = 6 \pi \sum_{n=1}^{\infty} \dfrac{ A_{n}^{} \Omega^{2}_{n}}{\sinh \left(\Omega_{n}\pi\right)} \cos  \left( \Omega_{n} \tau_{0} \right) - \dfrac{12\zeta}{5}.
\end{align}
Due to the summation, it is not possible to formulate an explicit condition for simple zeros, but instead perform a calculation numerically. For the Bessel-Fubini and the Fay solutions it is simple to show numerically that the Mel'nikov integral has simple zeroes for larger~$A_{n}$ than in the case of linear wave propagation. A Poincar\'e section can be constructed which is topologically conjugate to a Bernoulli shift map on two symbols - in effect the likelihood of a $R_{\mathrm{max}}$ being greater or smaller than the previous cycle is as random as the toss of a coin. Thus, for sufficiently small~$\varepsilon$ chaotic bubble oscillations will occur in the vicinity of $\boldsymbol{\gamma}$. From the simplified normal form, the effect of nonlinear wave propagation implies that a cascade to chaos and unbounded bubble growth will occur at higher driving pressures~$P$ or larger initial radii~$R_{0}$ than for linear wave propagation. Note that the classical first order Mel'nikov integral will typically overestimate the threshold by not including higher order contributions. 
\par
For the Bessel-Fubini solution as ${r_{s} \rightarrow 0}$ the linear result is recovered and no summation is required, that is as ${r_{s} \rightarrow 0}$ so ${P_{1} \rightarrow P}$, ${P_{j} \rightarrow 0}$ for all ${j=2,3, \ldots}$. As the distance towards the shock decreases, i.e.~$r_{s}$ increases from zero, so a higher $P$ or larger $R_{0}$ is necessary in order to have simple zeroes. For the Fay solution higher pressures are required further from the shock.  For low Gol'dberg numbers, then higher pressures are required than for materials with high Gol'dberg numbers.
\par
As ${\Omega \rightarrow \infty}$ then ${\mathcal{M}_{h}^{\left(1\right)} \rightarrow 12 \zeta \slash 5}$ and higher order contributions vanish, so that the stable and unstable manifolds stay disjoint. However, as ${\Omega \rightarrow 0}$ then ${\mathcal{M}_{h}^{\left(1\right)} \rightarrow 12 \zeta \slash 5}$, but now second order terms will affect the threshold criteria~\cite{lenci2004hmf}. Indeed, numerical analysis suggests that higher order contributions must be taken into account in this regime. Furthermore, the authors~\cite{lenci2004hmf} state that higher-order Mel'nikov analysis is always necessary when considering nonlinear wave propagation due to the interactions of the differing harmonic excitations.   
\par
It is simple to show that in the absence of (nondimensional) viscous damping, i.e.~${\zeta=0}$, that generically the Mel'nikov integral will have simple zeros for all (non-zero) parameter values. The effect of viscosity actually reduces the likelihood of violent cavitation, a result also found by Szeri~\cite{szeri1991oco}. No such analysis has yet been performed in the visco-elastic case but it is currently under investigation.
\par
\begin{figure}
\begin{center}
\subfigure[][]{
\begin{picture}(0,0)%
\includegraphics[scale=1.0]{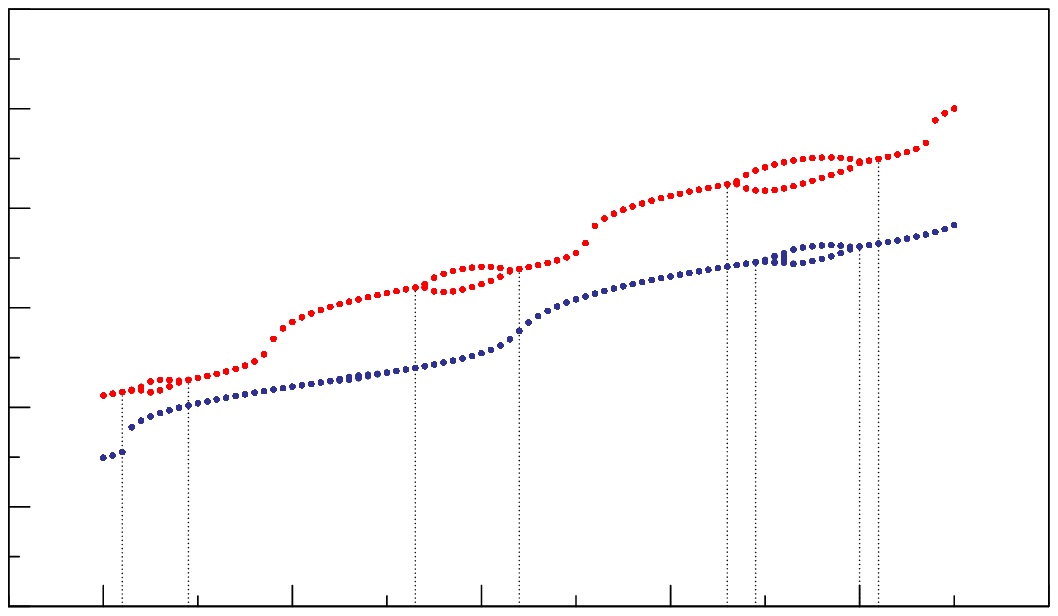} \label{fig:bif_contrast_R}%
\end{picture}%
\begingroup
\setlength{\unitlength}{0.0200bp}%
\begin{picture}(18000,10800)(0,0)%
  \put(1925,1650){\makebox(0,0)[r]{\strut{} 0}}%
  \put(1925,3083){\makebox(0,0)[r]{\strut{} 2}}%
  \put(1925,4517){\makebox(0,0)[r]{\strut{} 4}}%
  \put(1925,5950){\makebox(0,0)[r]{\strut{} 6}}%
  \put(1925,7383){\makebox(0,0)[r]{\strut{} 8}}%
  \put(1925,8817){\makebox(0,0)[r]{\strut{} 10}}%
  \put(1925,10250){\makebox(0,0)[r]{\strut{} 12}}%
  \put(3561,1100){\makebox(0,0){\strut{} 1.3}}%
  \put(6284,1100){\makebox(0,0){\strut{} 1.4}}%
  \put(9007,1100){\makebox(0,0){\strut{} 1.5}}%
  \put(11730,1100){\makebox(0,0){\strut{} 1.6}}%
  \put(14452,1100){\makebox(0,0){\strut{} 1.7}}%
  \put(17175,1100){\makebox(0,0){\strut{} 1.8}}%
  \put(550,5950){\rotatebox{90}{\makebox(0,0){\strut{}$\dfrac{R_{\mathrm{max}}-R_{0}}{R_{0}}$}}}%
  \put(9687,275){\makebox(0,0){\strut{}$P$[atm]}}%
\end{picture}%
\endgroup }
\subfigure[][]{
\begin{picture}(0,0)%
\includegraphics[scale=1.0]{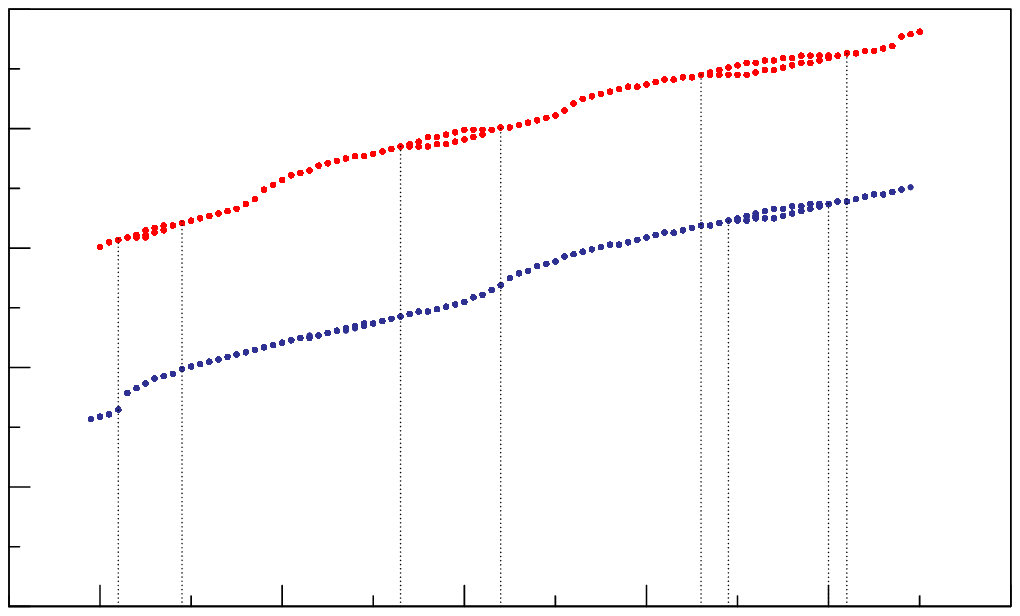} \label{fig:bif_contrast_eta}
\end{picture}%
\begingroup
\setlength{\unitlength}{0.0200bp}%
\begin{picture}(18000,10800)(0,0)%
  \put(2475,1650){\makebox(0,0)[r]{\strut{} 0.3}}%
  \put(2475,3370){\makebox(0,0)[r]{\strut{} 0.35}}%
  \put(2475,5090){\makebox(0,0)[r]{\strut{} 0.4}}%
  \put(2475,6810){\makebox(0,0)[r]{\strut{} 0.45}}%
  \put(2475,8530){\makebox(0,0)[r]{\strut{} 0.5}}%
  \put(2475,10250){\makebox(0,0)[r]{\strut{} 0.55}}%
  \put(4061,1100){\makebox(0,0){\strut{} 1.3}}%
  \put(6684,1100){\makebox(0,0){\strut{} 1.4}}%
  \put(9307,1100){\makebox(0,0){\strut{} 1.5}}%
  \put(11930,1100){\makebox(0,0){\strut{} 1.6}}%
  \put(14552,1100){\makebox(0,0){\strut{} 1.7}}%
  \put(17175,1100){\makebox(0,0){\strut{} 1.8}}%
  \put(550,5950){\rotatebox{90}{\makebox(0,0){\strut{}$\xi$}}}%
  \put(9962,275){\makebox(0,0){\strut{}$P$[atm]}}%
\end{picture}%
\endgroup }
\caption[]{Bifurcation diagram for a bubble determined by the Rayleigh-Plesset equation~\eqref{eq:RP} with parameters given in figure~\ref{fig:contrast1} but initial radius ${R_{0}=0.9\mu}$m. The figure shows the normalised maximum amplitudes for fifty cycles after fifty cycles in order to disregard transient behaviour. The first ten terms of the Bessel-Fubini solution~\eqref{eq:bessel_fubini} were computed. The radii from linear wave propagation are shown in red, those from nonlinear wave propagation shown in blue.  \label{fig:bif_contrast} }
\end{center}
\end{figure}
\begin{figure}
\begin{center}
\begin{picture}(0,0)%
\includegraphics{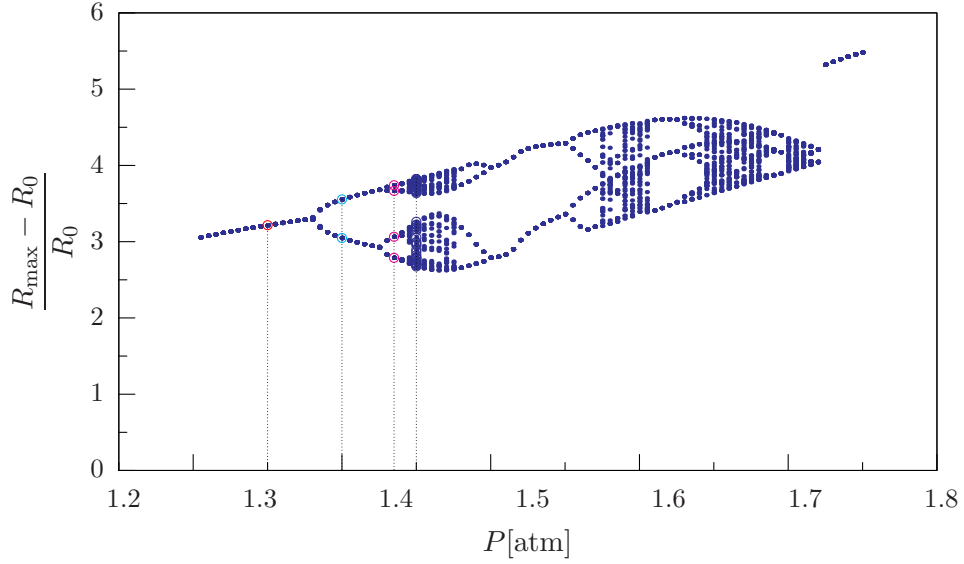}%
\end{picture}%
\begingroup
\setlength{\unitlength}{0.0200bp}%
\begin{picture}(18000,10800)(0,0)%
  \put(1650,1650){\makebox(0,0)[r]{\strut{} \small 0}}%
  \put(1650,3083){\makebox(0,0)[r]{\strut{} \small 1}}%
  \put(1650,4517){\makebox(0,0)[r]{\strut{} \small 2}}%
  \put(1650,5950){\makebox(0,0)[r]{\strut{} \small 3}}%
  \put(1650,7383){\makebox(0,0)[r]{\strut{} \small 4}}%
  \put(1650,8817){\makebox(0,0)[r]{\strut{} \small 5}}%
  \put(1650,10250){\makebox(0,0)[r]{\strut{} \small 6}}%
  \put(1925,1100){\makebox(0,0){\strut{} \small 1.2}}%
  \put(4467,1100){\makebox(0,0){\strut{} \small 1.3}}%
  \put(7008,1100){\makebox(0,0){\strut{} \small 1.4}}%
  \put(9550,1100){\makebox(0,0){\strut{} \small 1.5}}%
  \put(12092,1100){\makebox(0,0){\strut{} \small 1.6}}%
  \put(14633,1100){\makebox(0,0){\strut{} \small 1.7}}%
  \put(17175,1100){\makebox(0,0){\strut{} \small 1.8}}%
  \put(550,5950){\rotatebox{90}{\makebox(0,0){\strut{}$\dfrac{R_{\mathrm{max}}-R_{0}}{R_{0}}$}}}%
  \put(9550,275){\makebox(0,0){\strut{}$P$[atm]}}%
\end{picture}%
\endgroup
\caption[]{Bifurcation diagram showing the period doubling and the onset of quasi-periodic oscillations as the magnitude of the forcing pressure of the Bessel-Fubini solution~\eqref{eq:bessel_fubini} is varied for the Rayleigh-Plesset equation~\eqref{eq:RP}. The parameters are the same as figure~\ref{fig:bif_contrast} except that now ${R_{0}=1.4\mu\mathrm{m}}$. Profiles of the radius time curves at the marked pressures are illustrated in figure~\ref{fig:bif}. \label{fig:bif1} } 
\end{center}
\end{figure}

The bifurcation diagrams presented in figures~\ref{fig:bif_contrast} and~\ref{fig:bif1} were computed for the Rayleigh-Plesset equation~\eqref{eq:RP} with an adaptive, explicit Runge-Kutta method of order $(8,5)$ due to Dormand and Prince~\cite{hairer1993sod}. In order to ignore transient behaviour the first fifty cycles were disregarded and the next fifty cycles plotted. Figure~\ref{fig:bif_contrast} generalises the inferences from the radius-time profiles displayed in figure~\ref{fig:contrast1}. Subfigure~\ref{fig:bif_contrast_R} illustrates that the maximum amplitude of the bubbles under forcing from nonlinear wave propagation is smaller than the maximum amplitude of bubbles under forcing from linear wave propagation. By defining~$\xi$ as the phase of the minimum radius ${R_{\mathrm{min}}=R\left(t_{\mathrm{min}}\right)}$ per acoustic cycle~\cite{simon2002pot} 
\begin{equation}
\xi = \left( t_{\mathrm{min}} - t \right) \Omega \nonumber
\end{equation} 
subfigure~\ref{fig:bif_contrast_eta} illustrates that inertial cavitation occurs earlier in each acoustic cycle for nonlinear waves than for linear waves. Both bifurcation diagrams in figure~\ref{fig:bif_contrast} show the same regions of developing instability as both solution measures determine Poincar\'e sections tangential to the motion of the cavity, i.e.\ when~${\dot{R}=0}$.
\par
\begin{figure}
\begin{center}
\subfigure[][$P=1.300$atm]{
\begin{picture}(0,0)%
\includegraphics[scale=0.50]{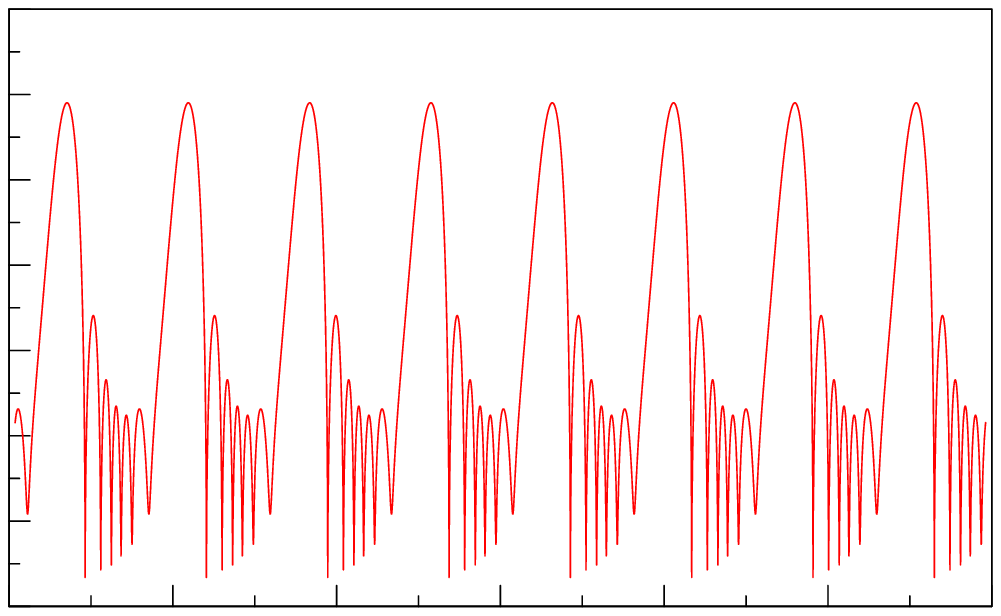}\label{fig:bif1350}%
\end{picture}%
\begingroup
\setlength{\unitlength}{0.0100bp}%
\begin{picture}(18000,10800)(0,0)%
  \put(2750,1650){\makebox(0,0)[r]{\strut{} \tiny 0}}%
  \put(2750,2879){\makebox(0,0)[r]{\strut{} \tiny 1e-5}}%
  \put(2750,4107){\makebox(0,0)[r]{\strut{} \tiny 2e-5}}%
  \put(2750,5336){\makebox(0,0)[r]{\strut{} \tiny 3e-5}}%
  \put(2750,6564){\makebox(0,0)[r]{\strut{} \tiny 4e-5}}%
  \put(2750,7793){\makebox(0,0)[r]{\strut{} \tiny 5e-5}}%
  \put(2750,9021){\makebox(0,0)[r]{\strut{} \tiny 6e-5}}%
  \put(2750,10250){\makebox(0,0)[r]{\strut{} \tiny 7e-5}}%
  \put(3025,1100){\makebox(0,0){\strut{} \tiny 0.00185}}%
  \put(5383,1100){\makebox(0,0){\strut{} \tiny 0.0019}}%
  \put(7742,1100){\makebox(0,0){\strut{} \tiny 0.00195}}%
  \put(10100,1100){\makebox(0,0){\strut{} \tiny 0.002}}%
  \put(12458,1100){\makebox(0,0){\strut{} \tiny 0.00205}}%
  \put(14817,1100){\makebox(0,0){\strut{} \tiny 0.0021}}%
  \put(17175,1100){\makebox(0,0){\strut{} \tiny 0.00215}}%
  \put(550,5950){\rotatebox{90}{\makebox(0,0){\strut{}$R$[m]}}}%
\put(10100,75){\makebox(0,0){\strut{}$t$[s]}}%
\end{picture}%
\endgroup }
\subfigure[][$P=1.350$atm]{
\begin{picture}(0,0)%
\includegraphics[scale=0.50]{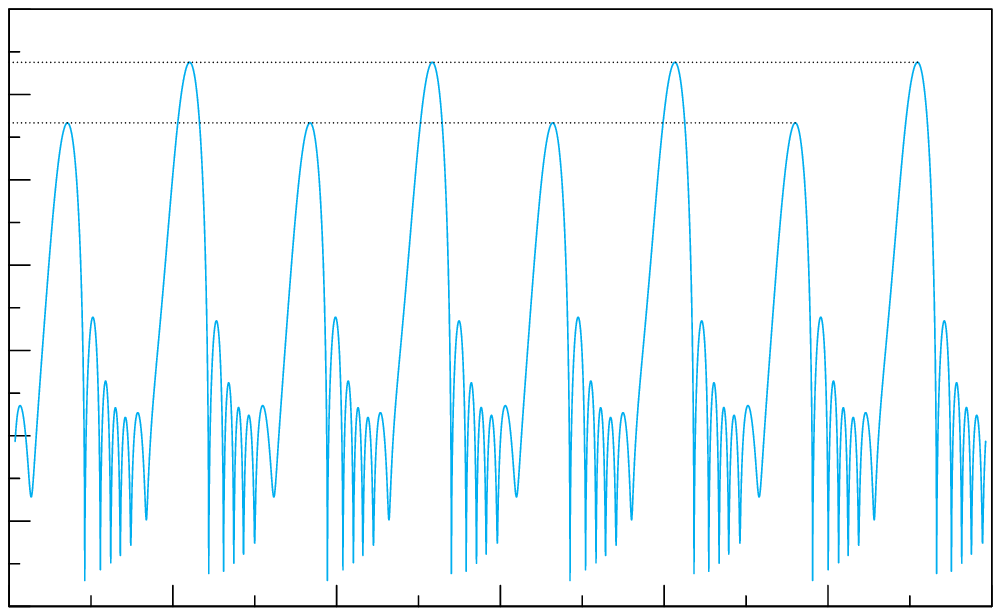}\label{fig:bif1400}%
\end{picture}%
\begingroup
\setlength{\unitlength}{0.0100bp}%
\begin{picture}(18000,10800)(0,0)%
  \put(2750,1650){\makebox(0,0)[r]{\strut{} \tiny 0}}%
  \put(2750,2879){\makebox(0,0)[r]{\strut{} \tiny 1e-5}}%
  \put(2750,4107){\makebox(0,0)[r]{\strut{} \tiny 2e-5}}%
  \put(2750,5336){\makebox(0,0)[r]{\strut{} \tiny 3e-5}}%
  \put(2750,6564){\makebox(0,0)[r]{\strut{} \tiny 4e-5}}%
  \put(2750,7793){\makebox(0,0)[r]{\strut{} \tiny 5e-5}}%
  \put(2750,9021){\makebox(0,0)[r]{\strut{} \tiny 6e-5}}%
  \put(2750,10250){\makebox(0,0)[r]{\strut{} \tiny 7e-5}}%
  \put(3025,1100){\makebox(0,0){\strut{} \tiny 0.00185}}%
  \put(5383,1100){\makebox(0,0){\strut{} \tiny 0.0019}}%
  \put(7742,1100){\makebox(0,0){\strut{} \tiny 0.00195}}%
  \put(10100,1100){\makebox(0,0){\strut{} \tiny 0.002}}%
  \put(12458,1100){\makebox(0,0){\strut{} \tiny 0.00205}}%
  \put(14817,1100){\makebox(0,0){\strut{} \tiny 0.0021}}%
  \put(17175,1100){\makebox(0,0){\strut{} \tiny 0.00215}}%
  \put(550,5950){\rotatebox{90}{\makebox(0,0){\strut{}$R$[m]}}}%
  \put(10100,75){\makebox(0,0){\strut{}$t$[s]}}%
\end{picture}%
\endgroup }
\subfigure[][$P=1.375$atm]{
\begin{picture}(0,0)%
\includegraphics[scale=0.50]{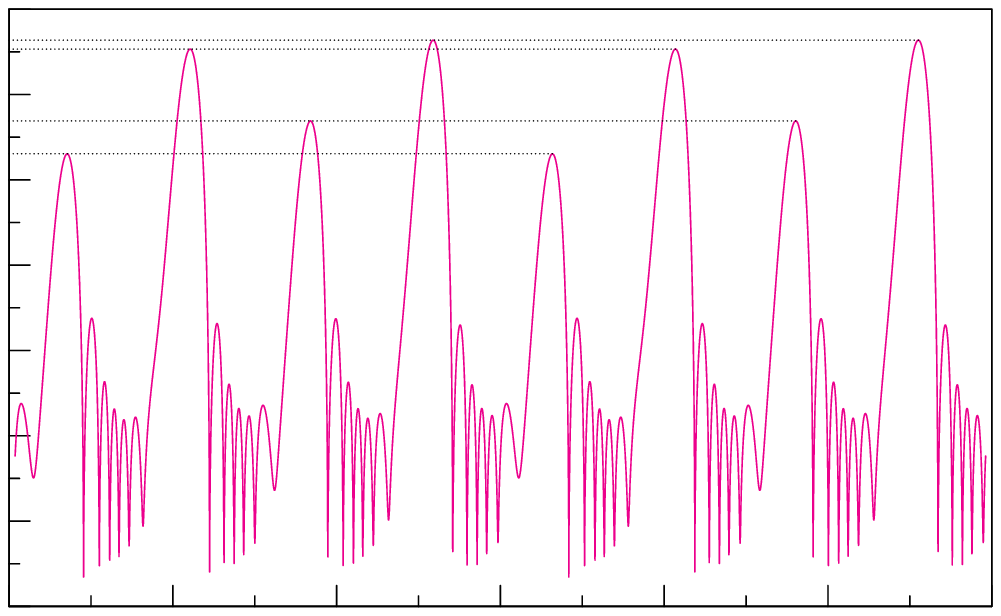}\label{fig:bif1435}%
\end{picture}%
\begingroup
\setlength{\unitlength}{0.0100bp}%
\begin{picture}(18000,10800)(0,0)%
  \put(2750,1650){\makebox(0,0)[r]{\strut{} \tiny 0}}%
  \put(2750,2879){\makebox(0,0)[r]{\strut{} \tiny 1e-5}}%
  \put(2750,4107){\makebox(0,0)[r]{\strut{} \tiny 2e-5}}%
  \put(2750,5336){\makebox(0,0)[r]{\strut{} \tiny 3e-5}}%
  \put(2750,6564){\makebox(0,0)[r]{\strut{} \tiny 4e-5}}%
  \put(2750,7793){\makebox(0,0)[r]{\strut{} \tiny 5e-5}}%
  \put(2750,9021){\makebox(0,0)[r]{\strut{} \tiny 6e-5}}%
  \put(2750,10250){\makebox(0,0)[r]{\strut{} \tiny 7e-5}}%
  \put(3025,1100){\makebox(0,0){\strut{} \tiny 0.00185}}%
  \put(5383,1100){\makebox(0,0){\strut{} \tiny 0.0019}}%
  \put(7742,1100){\makebox(0,0){\strut{} \tiny 0.00195}}%
  \put(10100,1100){\makebox(0,0){\strut{} \tiny 0.002}}%
  \put(12458,1100){\makebox(0,0){\strut{} \tiny 0.00205}}%
  \put(14817,1100){\makebox(0,0){\strut{} \tiny 0.0021}}%
  \put(17175,1100){\makebox(0,0){\strut{} \tiny 0.00215}}%
  \put(550,5950){\rotatebox{90}{\makebox(0,0){\strut{}$R$[m]}}}%
\put(10100,75){\makebox(0,0){\strut{}$t$[s]}}%
\end{picture}%
\endgroup }
\subfigure[][$P=1.400$atm]{
\begin{picture}(0,0)%
\includegraphics[scale=0.50]{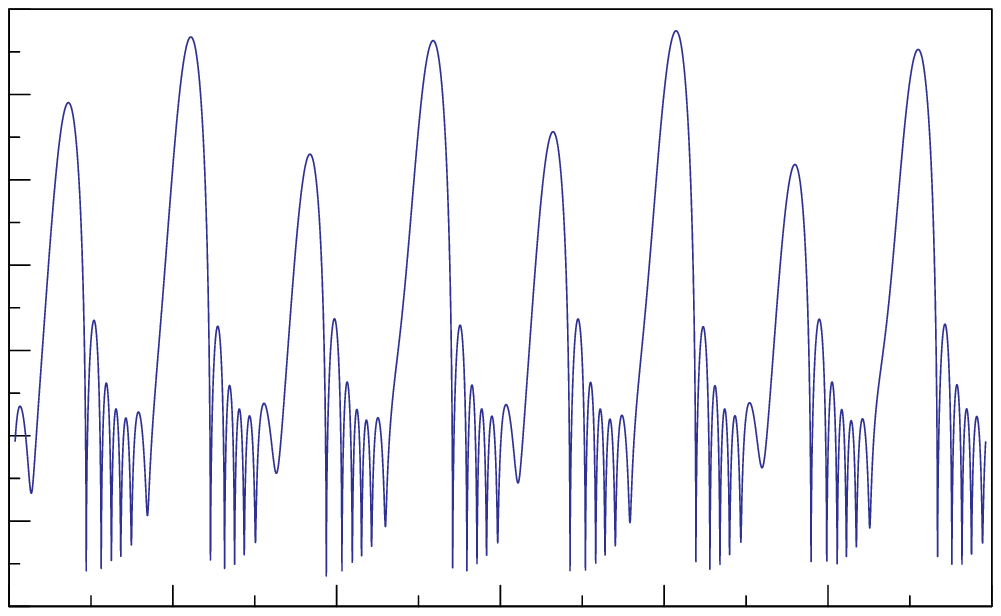}\label{fig:bif1450}%
\end{picture}%
\begingroup
\setlength{\unitlength}{0.0100bp}%
\begin{picture}(18000,10800)(0,0)%
  \put(2750,1650){\makebox(0,0)[r]{\strut{} \tiny 0}}%
  \put(2750,2879){\makebox(0,0)[r]{\strut{} \tiny 1e-5}}%
  \put(2750,4107){\makebox(0,0)[r]{\strut{} \tiny 2e-5}}%
  \put(2750,5336){\makebox(0,0)[r]{\strut{} \tiny 3e-5}}%
  \put(2750,6564){\makebox(0,0)[r]{\strut{} \tiny 4e-5}}%
  \put(2750,7793){\makebox(0,0)[r]{\strut{} \tiny 5e-5}}%
  \put(2750,9021){\makebox(0,0)[r]{\strut{} \tiny 6e-5}}%
  \put(2750,10250){\makebox(0,0)[r]{\strut{} \tiny 7e-5}}%
  \put(3025,1100){\makebox(0,0){\strut{} \tiny 0.00185}}%
  \put(5383,1100){\makebox(0,0){\strut{} \tiny 0.0019}}%
  \put(7742,1100){\makebox(0,0){\strut{} \tiny 0.00195}}%
  \put(10100,1100){\makebox(0,0){\strut{} \tiny 0.002}}%
  \put(12458,1100){\makebox(0,0){\strut{} \tiny 0.00205}}%
  \put(14817,1100){\makebox(0,0){\strut{} \tiny 0.0021}}%
  \put(17175,1100){\makebox(0,0){\strut{} \tiny 0.00215}}%
  \put(550,5950){\rotatebox{90}{\makebox(0,0){\strut{}$R$[m]}}}%
  \put(10100,75){\makebox(0,0){\strut{}$t$[s]}}%
\end{picture}%
\endgroup  }
\caption[short]{Period-doubling cascade to chaos as illustrated by the four subfigures marked on the bifurcation diagram~\ref{fig:bif1}. Subfigure~\ref{fig:bif1350} shows a stable one-period radius-time profile at ${P=1.30}$atm, which undergoes a period doubling bifurcation so that at ${P=1.35}$atm in subfigure~\ref{fig:bif1400} the profile is two-period. For ${P=1.375}$atm as subfigure~\ref{fig:bif1435} is four period before subfigure~\ref{fig:bif1450} shows a chaotic profile at ${P=1.40}$atm.} 
\label{fig:bif}
\end{center}
\end{figure}

Figure~\ref{fig:bif1} shows a bifurcation diagram beyond the critical threshold and figure~\ref{fig:bif} then shows a selection of associated radius-time profiles illustrating the cascade to chaos through a period doubling bifurcation. Subfigure~\ref{fig:bif1350} shows a stable one-period radius-time profile at ${P=1.30}$atm, which undergoes a period doubling bifurcation so that at ${P=1.35}$atm in subfigure~\ref{fig:bif1400} the profile is two-periodic. For ${P=1.375}$atm as subfigure~\ref{fig:bif1435} illustrates the radius-time profile is four-periodic. By ${P=1.40}$atm the pressure has passed an accumulation point of period doubling bifurcations and as subfigure~\ref{fig:bif1450} illustrates the bubble oscillations are chaotic. Note that the pressure intervals between periodic cycles decreases as the system approaches chaotic oscillations, characteristic of the cascade to chaos.  From a clinical perspective, when the duty cycle maybe in the order of seconds rather than micro seconds, accurate predictions for the heat deposition on tumour sites over all but the shortest time scales can not be inferred from cavitation activity when the driving pressure is beyond the threshold. 
\par
Bifurcation diagrams of linear and nonlinear wave propagation at equal driving pressures far beyond the threshold are significantly different, with differing regions of stability, indicating that beyond the threshold predictions based on linear wave propagation will be inaccurate.  
\section{Conclusion} \label{sec:conclusions}
In this paper the effect of nonlinear wave propagation has been investigated numerically and analytically. The effect of nonlinear wave propagation redistributes energy from the primary harmonic to higher harmonics. From this two significant conclusions can be drawn; firstly the maximum bubble radius is reduced. This is clearly illustrated by the bifurcation diagram~\ref{fig:bif_contrast_R}. Thus the likelihood of shape instability is reduced. Secondly, inertial cavitation occurs earlier in each cycle, again as illustrated by the bifurcation diagram~\ref{fig:bif_contrast_eta}. The earlier onset of collapse allows for the bubble to return to an equilibrium radius and for inertial cavitation to reoccur. 
\par
Amongst the many threshold criteria applied to cavitation, it is worthwhile emphasising, although the Mel'nikov criteria does not exactly correspond to the escape boundary whereby bubbles increase without bound~\cite{thompson1989cpt}, it initiates the penetration of escaping tongues into the safe basin and therefore constitutes the first event in a well known sequence of complicated events which leads to unbounded bubble growth. Thus the Mel'nikov criteria provides a lower bound threshold value. In the context of therapeutic applications, where safety is paramount, such a criteria is of great value.
\par
In therapeutic applications a cloud of bubbles will exist, comprised of many thousands of bubbles of differing equilibrium radii and resonance frequencies. Thus it could be assumed that stability criterion for a single bubble is of limited use. However, the stability of a single bubble does provide insight into the stability of entire bubble clouds: experimental evidence~\cite{lauterborn1987hop} suggests that entire bubble clouds undergo period doubling cascade to chaos, not just the subset of bubbles which satisfy the Mel'nikov criterion. It is believed that the interaction between the bubbles results in an averaged behaviour. Indeed, numerical and experimental calculations of the dimension of the attractor in phase space are remarkably low, between 2 and 2.5, indicating that the number of relevant degrees of freedom in the system is also low~\cite{lauterborn1986eld,holt1994cs}. 
\par
In many applications of therapeutic ultrasound the tissue will have distributed inhomogeneities, leading to dissipation through diffraction, but it is conjectured that material nonlinearity of is greater significance than material inhomogeneity~\cite{duck1990ppt}, although this subject is currently under investigation. 
\ack
This work was supported under grant EP/F025750/1 and is gratefully acknowledged. The authors would like to thank S.~Martynov, G.~Vilensky and P. G\'{e}lat for insightful comments and discussions.
\section*{References}
\bibliographystyle{iopart-num}
\bibliography{arxiv} 
\end{document}